\documentclass[conference]{IEEEtran}
\IEEEoverridecommandlockouts
\usepackage{cite}
\usepackage{amsmath,amssymb,amsfonts}
\usepackage{graphicx}
\usepackage{booktabs}
\usepackage{enumitem}
\usepackage{xurl}
\usepackage{hyperref}
\usepackage{caption}
\usepackage{tikz}
\usepackage{float}
\usetikzlibrary{positioning, arrows.meta, shapes.geometric, calc, patterns}

\graphicspath{{Content/}}
\hypersetup{colorlinks=true, linkcolor=blue, urlcolor=cyan, citecolor=blue}

\begin{document}
\title{Microbenchmark-Driven Analytical Performance Modeling Across Modern GPU Architectures}

\author{
\IEEEauthorblockN{Aaron Jarmusch}
\IEEEauthorblockA{\textit{dept. Computer \& Information Sciences} \\
\textit{University of Delaware}\\
Newark, Delaware, USA}
\and
\IEEEauthorblockN{Sunita Chandrasekaran}
\IEEEauthorblockA{\textit{dept. Computer \& Information Sciences} \\
\textit{University of Delaware}\\
Newark, Delaware, USA}
}

\maketitle

\begin{abstract}
Rapidly evolving GPU architectures featuring complex memory hierarchies, matrix units, and varied precision formats continue to widen the gap between theoretical peaks and achievable performance. We design and develop analytical performance models for NVIDIA Blackwell (B200) and AMD CDNA3 (MI300A) grounded in systematic microbenchmark characterization. For Blackwell, the model captures Tensor Memory (TMEM), asynchronous bulk copy (TMA), and 5th-generation tensor cores; for CDNA3, the model captures Infinity Cache hierarchy, VGPR constraints, and occupancy. Validation yields 1.31\% MAE on B200 (21 kernels) and $\sim$0.09\% on MI300A (27 kernels), while naive roofline baselines exceed 95\% error on the same kernels. We further validate the models using Rodinia~3.1 and SPEChpc 2021 Tiny.The models are updated with HBM bandwidth, capacity, and cache parameters and applied to H200 (Hopper) and MI250X (CDNA2), indicating no major restructuring of the models are needed. All models and benchmarks will be released as open-source upon acceptance.
\end{abstract}


\section{Introduction}
\label{sec:introduction}

Modern HPC and AI systems increasingly rely on GPU accelerators that expose distinct execution primitives across vendors and generations (For example, Tensor Memory (TMEM) and asynchronous bulk copy (TMA) on NVIDIA Blackwell, wavefront scheduling and Infinity Cache on AMD CDNA), and whose sustained performance diverges significantly from datasheet peaks~\cite{evolutionGPUs,kurowski2011parallel,koilia2024hardware}. Every 18--24 months a new architecture is introduced with new memory subsystems, matrix execution units, and precision formats. Understanding the gap between theoretical peak and achievable efficiency requires architecture-aware models that capture memory hierarchy behavior, matrix unit utilization, and occupancy constraints, not just bandwidth-limited roofline bounds.

This paper presents systematic microbenchmark-driven stage-centric analytical models for two current-generation GPU accelerators: NVIDIA Blackwell (B200)~\cite{NVIDIA2024_Blackwell} and AMD CDNA3 (MI300A)~\cite{AMD_CDNA3_WhitePaper_2023}. Stage-centric implies the different  stages in a GPU compilation. Our models are build on prior microbenchmark characterizations of both architectures~\cite{jarmusch2025blackwell,jarmusch2026mi300a}. We characterize each architecture through targeted low-level benchmarks, derive model parameters directly from measurements, and validate against both a microbenchmark suite and two HPC application benchmarks (Rodinia~3.1~\cite{luhnen2024benchmarking} and SPEChpc 2021 Tiny~\cite{gilman2020demystifying}). The models are interpretable, parameterized by measured hardware values, and accurate to within 1--5\% MAE, compared to over 95\% error for naive roofline. We then apply the same methodology to the prior generation of each vendor (NVIDIA Hopper H200 and AMD CDNA2 MI250X), demonstrating that both ports required only hardware parameter updates with no model re-derivation.

We make the following novel contributions:
\begin{itemize}
    \item Construct and validate a stage-centric analytical model for NVIDIA Blackwell (B200) capturing TMA, TMEM, 5th-generation tensor cores, and the 2-SM cooperative execution model; to our knowledge the first validated execution-time model for this architecture
    \item Construct and validate stage-centric analytical model for AMD CDNA3 (MI300A) accounting for Infinity Cache hierarchy, VGPR register pressure, and occupancy-driven tile selection
    \item Validate both models against Rodinia~3.1 and SPEChpc 2021 Tiny on B200 and MI300A, with naive roofline baselines showing $>$95\% error
    \item By updating parameters of our analytical models, they can be applied to H200 and MI250X. 
\end{itemize}

\textbf{Paper organization.} Section~\ref{sec:related-work} positions our work against prior analytical and ML-based models. Section~\ref{sec:overview} summarizes Blackwell and CDNA architectures and their implications for modeling. Section~\ref{sec:execution-model} presents the stage-centric model for Blackwell and the wavefront-centric model for AMD CDNA. Section~\ref{sec:results} describes the microbenchmark validation setup, results, and cross-platform comparison with naive roofline baselines. Section~\ref{sec:observation} discusses systematic bias, limitations, architectural insights, and portability. Section~\ref{sec:conclusion} presents the conclusion.

\section{Related Work}
\label{sec:related-work}
Related work spans roofline modeling, analytical and simulation-based GPU performance prediction, and microarchitecture characterization.

\subsection{Roofline and Analytical Models}
The roofline model~\cite{williams2009roofline} relates arithmetic intensity to attainable performance, identifying memory-bound vs.\ compute-bound regimes. Cache-aware extensions~\cite{ilic2014cacheawareroofline,ofenbeck2014applyingroofline} and AMD instruction-roofline variants~\cite{Leinhauser2022_RooflineAMD} improve diagnosis, but all remain single-axis intensity plots that cannot capture per-stage pipeline overlap, dedicated matrix paths, or TMEM residency.

\textbf{Why roofline gives $>$95\% error on modern kernels.}
The naive bound $T_{\text{roofline}} = \max(\text{FLOPs}/P_{\text{peak}},\, \text{bytes}/B_{\text{HBM}})$ fails for three compounding reasons: (1)~datasheet peaks overstate achievable throughput (B200 sustained tensor-core throughput is 1,100--1,400~TFLOPS vs.\ 2,250 datasheet; sustained HBM is 6.8--7.1~TB/s vs.\ 8.0), making $T_{\text{roofline}}$ too small by $1.5\text{--}2\times$; (2)~roofline ignores serialized pipeline stages (on Blackwell, TMA$\to$TMEM$\to$Tensor Core stages add latency that max() cannot represent); (3)~roofline uses a single bandwidth figure, missing MI300A's 256~MB Infinity Cache (17.2~TB/s vs.\ 5.3~TB/s HBM). These compound multiplicatively to $>$90\% error (Table~\ref{tab:roofline_baseline}).

Hong and Kim~\cite{hong2009analytical} established the MWP/CWP framework for bounding parallelism on early NVIDIA GPUs; we retain that occupancy intuition but add Blackwell-specific TMEM/TMA/sync terms. GCoM~\cite{Lee2022_GCoM} provides a detailed Ampere-era core model; we target newer architectures. GPUMech~\cite{huang2014gpumech} and MDM~\cite{wang2020mdm} address interval analysis and memory divergence respectively, neither targeting tensor cores or TMEM.

\subsection{Simulation, Characterization, and ML Approaches}
Accel-Sim~\cite{Khairy2020_AccelSim} offers cycle-accurate simulation but depends on architecture definition files often unavailable for the newest accelerators. Analytical models trade accuracy on irregular control flow for interpretability and negligible evaluation cost.

Key characterization work: Wong et al.~\cite{wong2010demystifying} established GPU microbenchmarking methodology; Jia et al.~\cite{jia2018dissecting} dissected Volta; Luo et al.~\cite{luo2025hopper} characterized Hopper TMA/wgmma/DSM (basis for our H200 calibration). On AMD, Wahlgren et al.~\cite{wahlgren2025dissectingcpugpuunifiedphysical} studied MI300A unified memory and Schieffer et al.~\cite{schieffer2024amd_matrix_cores} characterized matrix-core performance. Jarmusch et al.~\cite{jarmusch2025blackwell} provided the first Blackwell (B200) microbenchmark characterization; Jarmusch et al.~\cite{jarmusch2026mi300a} characterized MI300A FP8 matrix cores and ACE concurrency. Our analytical models build directly on these measurements. Chowdhury et al.~\cite{chowdhury2020tcu} modeled tensor core units; AMALI~\cite{cao2025amali} targets LLM inference; Fasi et al.~\cite{Fasi2021} analyze tensor-core numerics. None deliver full-kernel analytical models with reported MAE across vendors.

\subsection{Positioning}
Our work is, to our knowledge, the first to (a)~provide validated analytical execution-time models for Blackwell (B200) and CDNA3 (MI300A), (b)~update parameters of our analytical models and apply to H200 and MI250X, and (c)~report cross-vendor MAE under a shared protocol. Prior models require full re-derivation or simulator updates when new hardware ships; we close this gap by tying every coefficient to a microbenchmark. Table~\ref{tab:related-work} summarizes.

\begin{table*}[t]
\centering
\caption{Related work vs.\ this paper: coverage of modern features, cross-vendor validation, and porting methodology.}
\footnotesize
\begin{tabular}{@{}lccll@{}}
\toprule
\textbf{Work} & \textbf{Arch.} & \textbf{TMEM/TMA/Decomp.} & \textbf{Cross-vendor} & \textbf{Validated MAE} \\
\midrule
Hong et al.~\cite{hong2009analytical} & NVIDIA & -- & No & -- \\
GCoM~\cite{Lee2022_GCoM} & NVIDIA & -- & No & Yes \\
Chowdhury et al.~\cite{chowdhury2020tcu} & Tensor Cores & TC only & No & -- \\
Wahlgren/Schieffer~\cite{wahlgren2025dissectingcpugpuunifiedphysical,schieffer2024amd_matrix_cores} & AMD MI300A & -- & No & Partial \\
\textbf{This work} & Blackwell, CDNA~3 & Yes & Yes & 1.3\% (B200), $\sim$0.09\% (MI300A) \\
\bottomrule
\end{tabular}
\label{tab:related-work}
\end{table*}

\section{GPU Architecture Overview}
\label{sec:overview}
\begin{figure}[t]
\centering
\includegraphics[width=1\linewidth]{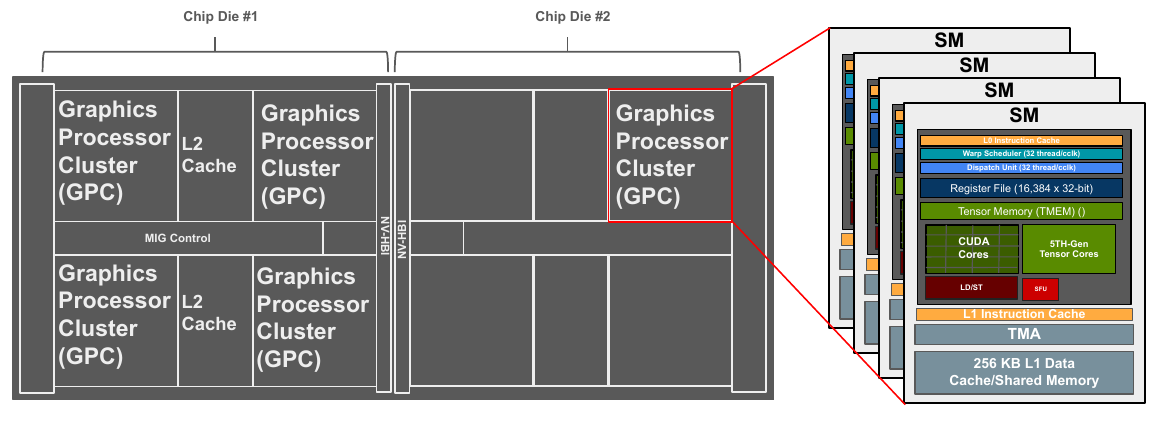}
\caption{NVIDIA Blackwell architecture: dual-die, TMEM, and SM structure.}
\label{fig:b200-arch}
\vspace{-1em}
\end{figure}

We focus on the NVIDIA Blackwell B200~\cite{NVIDIA2024_Blackwell}  and AMD CDNA 3 MI300A~\cite{AMD_CDNA3_WhitePaper_2023}; two state-of-the-art architectures representing the leading edge of heterogeneous integration and matrix acceleration.

\textbf{NVIDIA Blackwell (B200).} Dual-die design (208B transistors) with NV-HBI (10~TB/s inter-die), unified 192~GB HBM3e and cache coherence~\cite{NVIDIA2024_Blackwell}. Fifth-gen Tensor Cores support FP4/FP6 with up to 9,000~TFLOPS FP4; Transformer Engine improves low-precision stability. \textbf{Tensor Memory (TMEM)} provides 256~KB/SM for tensor ops, reducing shared-memory contention. A \textbf{decompression engine} supports LZ4/Snappy/Deflate in the data path. The Tensor Memory Accelerator (TMA) handles hardware-managed async transfers; \texttt{tcgen05.mma} integrates TMEM and weight-stationary dataflows (Figure~\ref{fig:b200-arch}).

\begin{figure}[t]
\centering
\includegraphics[width=0.7\linewidth]{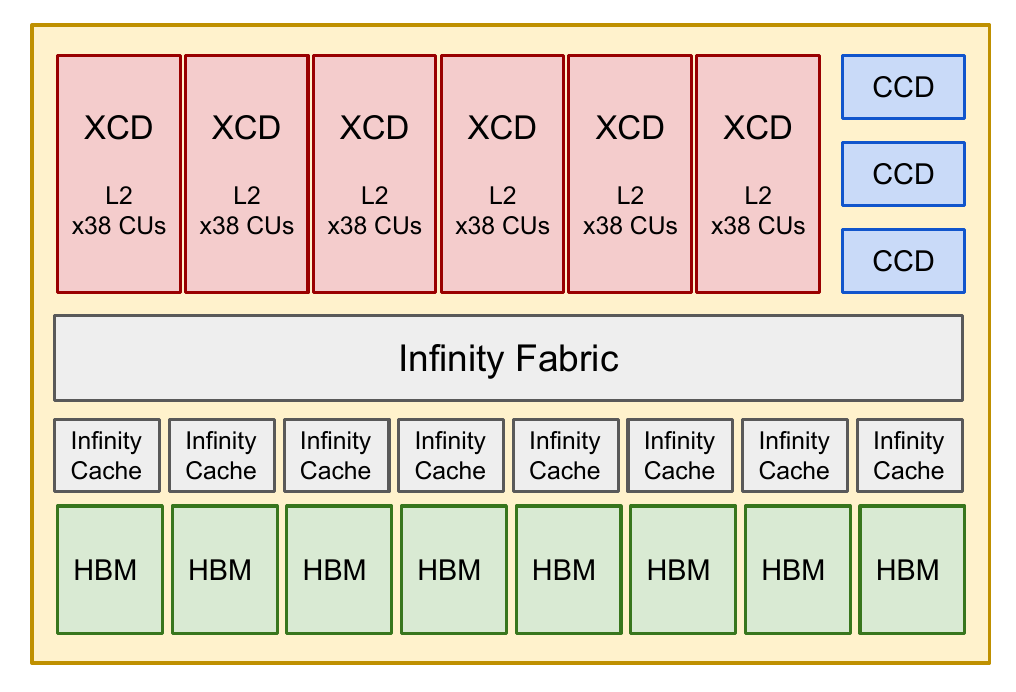}
\caption{MI300A APU: GPU/CPU chiplets and unified memory.}
\label{fig:amd-arch}
\vspace{-0.2in}
\end{figure}

\textbf{AMD MI300A.} First heterogeneous APU: six GPU chiplets (GCDs) and three Zen~4 CPU chiplets with 24 cores and \textbf{Unified Physical Memory (UPM)}—true hardware-coherent CPU–GPU sharing without explicit copies~\cite{AMD_CDNA3_WhitePaper_2023}. Matrix Cores support FP8 (1,307~TFLOPS) and strong FP64 (61.3~TFLOPS); 256~MB L2 (Infinity Cache) reduces latency for large working sets. Execution is wavefront-based (64 threads); 304~CUs (38 per XCD $\times$ 8 XCDs) with TF32 and 2:4 sparsity. Cross-XCD access incurs 50--100~ns NUMA-like penalties (Figure~\ref{fig:amd-arch}).

\textbf{Modeling implications.} Blackwell's explicit stages (TMA $\rightarrow$ TMEM $\rightarrow$ Tensor Core $\rightarrow$ Sync) allow stage-centric modeling with measurable latencies and bandwidths; roofline's single max() cannot represent their serialization. MI300A's overlap is implicit and occupancy-driven; accumulators live in VGPRs (vs.\ TMEM), creating tile-size vs.\ occupancy tradeoffs that a bandwidth roofline cannot capture. These structural differences are why a single roofline formulation gives $>$95\% error on both platforms (Table~\ref{tab:roofline_baseline}), and why the two architectures require distinct model frameworks rather than parameter substitution in a shared roofline. They also illustrate the principle of extensibility. If a future GPU introduces a distinct accumulation mechanism—for instance, dedicated register files for FP4 on a future CDNA4 or NV-HBI-aware TMEM on Rubin integration is a matter of identifying the most similar framework and adding the new term without the need to rebuild the model from the ground up. 
Table~\ref{tab:arch-params} makes this concrete: every parameter is either measured by a microbenchmark (bandwidth, throughput, max resident warps) or taken from the vendor datasheet (SM/CU count, cache sizes, capacity). Swapping in values for a new GPU updates the model without changing any formula.

% Table~\ref{tab:arch-params} summarizes the unified parameter set used in the analytical model for each architecture. All values are either from public documentation or from our microbenchmark suite (Section~\ref{sec:results}); no architecture-specific constants are hard-coded in the model so that calibration can be updated from new measurements.

\begin{table*}[t]
\centering
\caption{Architecture parameters for B200 and MI300A. Each value is from a microbenchmark or vendor datasheet (Source column). H200 and MI250X use the same model with their own values. FP64 roofline for SPEChpc uses 30.4~TFLOPS on MI300A.}
\footnotesize
\begin{tabular}{@{}llll@{}}
\toprule
\textbf{Parameter} & \textbf{B200} & \textbf{MI300A} & \textbf{Source} \\
\midrule
SMs / CUs & 176 & 304 & Datasheet \\
Warp / wavefront size & 32 & 64 & Datasheet \\
Max resident warps/wavefronts & 64 & 32 & Microbench / docs \\
HBM peak BW (TB/s) & 8.0 & 5.3 & Bandwidth microbench / datasheet \\
HBM capacity (GB) & 192 & 128 & Datasheet \\
L2 / LLC (MB) & 64 & 256 & Datasheet \\
TMEM / LDS (KB per SM/CU) & 256 & 64 & Datasheet \\
Tensor / MFMA peak (TFLOPS) & 2,250 (FP16), 4,500 (FP8) & 1,307 (FP8), 61.3 (FP64) & Throughput microbench / datasheet \\
\bottomrule
\end{tabular}
\label{tab:arch-params}
\end{table*}

\section{Analytical Model}
\label{sec:execution-model}
\begin{figure}[t]
\centering
\includegraphics[width=1\linewidth]{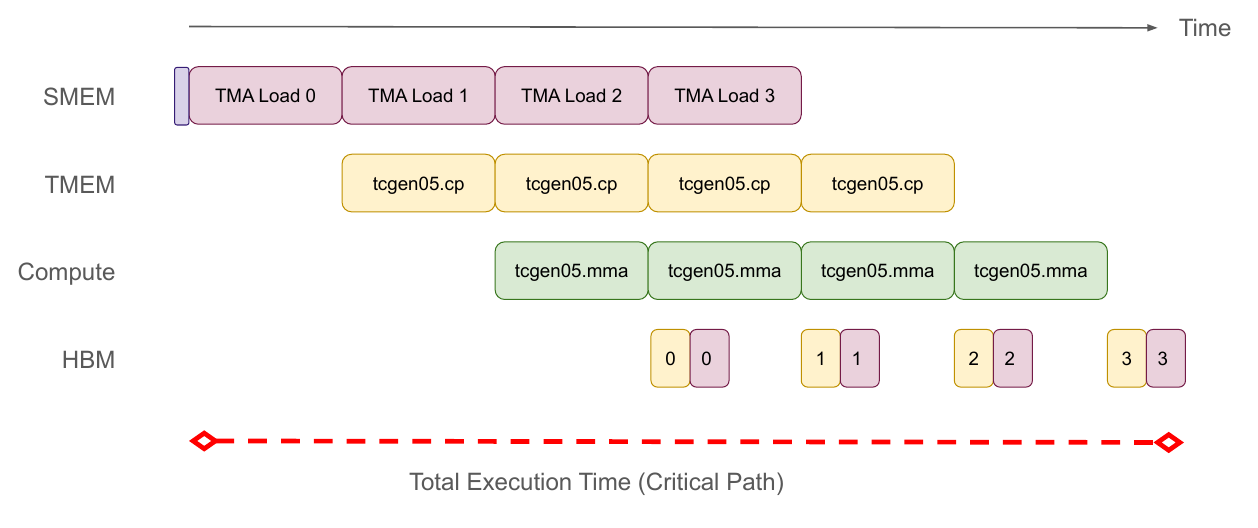}
\caption{Per-CTA execution: pipeline stages, overlap, and critical path.}
\label{fig:cta-exec-model}
\vspace{-0.2in}
\end{figure}

We adopt the Hong–Kim framework~\cite{hong2009analytical}: execution time is the maximum of compute and memory time plus overhead:
\begin{equation}
T_{\text{exec}} = \max(T_{\text{compute}}, T_{\text{memory}}) + T_{\text{overhead}}
\label{eq:hong-total-execution}
\end{equation}
$T_{\text{overhead}}$ covers barriers and kernel launch. Blackwell~\cite{NVIDIA2024_Blackwell} uses explicit stage-centric pipelines; MI300A~\cite{AMD_CDNA3_WhitePaper_2023} uses implicit wavefront-centric scheduling. Figure~\ref{fig:cta-exec-model} summarizes the per-CTA pipeline for Blackwell.

\subsection{NVIDIA Blackwell Model}

\subsubsection{Tensor Core Compute and TMEM}
TMEM (256~KB/SM) holds accumulators; access has measurable bandwidth. Per K-step tile iteration:
\begin{equation}
T_{\text{TMEM\_per\_tile}} = \frac{D_{\text{accum}}}{BW_{\text{TMEM\_read}}} + L_{\text{mma}} + \frac{D_{\text{accum}}}{BW_{\text{TMEM\_write}}}
\label{eq:tmem-single-tile-iteration}
\end{equation}
Exceeding 256~KB forces spill and hurts efficiency. Per-CTA compute time:
\begin{equation}
T_{\text{compute}} = \frac{2b_Mb_Nb_K}{R_{TC}^{SM} \times S_{\text{mode}}} + T_{\text{TMEM}} + T_{\text{TMEM\_mgmt}}
\label{eq:cta-tmem-compute-time}
\end{equation}
$R_{TC}^{SM}$ is tensor core throughput per SM; $S_{\text{mode}}$ accounts for 2-SM cooperation. Two TMEM strategies: (A) accumulators only in TMEM, inputs in SMEM—lower TMEM traffic; (B) A and accumulators in TMEM via \texttt{tcgen08.cp}—reduces SMEM pressure but increases TMEM bandwidth. The model covers both via $T_{\text{TMEM}}$ and measured bandwidths.

\subsubsection{Memory and TMA}
TMA multicast: $P$ participants share a tile; bytes per CTA: $bytes_{perCTA} = bytes(T)/P$. TMA time per CTA:
\begin{equation}
T_{tma} = L_{TMA} + \frac{bytes(T)}{P \times B_{TMA}}
\label{eq:percta-tma}
\end{equation}
$L_{TMA}$, $B_{TMA}$ from microbenchmarks; L2 hit rates strongly affect $B_{TMA}$. For different $P_A$, $P_B$ on A/B, sum or take max depending on overlap.

\subsubsection{Decompression and Synchronization}
Decompression (link vs.\ engine): $T_{DE\_load} = \max(D_{compressed}/BW_{link},\, D_{compressed}/R_{DE})$. With compression ratio $CR$ and efficiency $\eta_{DE}$:
\begin{equation}
T_{DE\_load} = \frac{D_{uncompressed}}{CR \times BW_{link} \times \eta_{DE}}
\label{eq:decompression-time}
\end{equation}
Sub-byte unpacking: $T_{decomp} = bytes_{comp}/R_{decomp} + L_{decomp\_setup}$; $R_{decomp}$ and overlap via $\alpha$. We use $\alpha \in [0.85, 0.95]$ from pipeline depth (double- to triple-buffering); sensitivity is discussed in Section~\ref{sec:results}. Sync per K-step: $T_{sync} = N_{bar} \times L_{mbar}$; $N_{bar}$ typically 1--2, $L_{mbar}$ from microbenchmarks.

\subsubsection{CTA Pairs (2-SM UMMA)}
CTA pairs on adjacent SMs share B via distributed shared memory (DSMEM). Traffic: $D_{2-CTA} = 2M_A + M_B$ (vs.\ $2(M_A+M_B)$), giving up to $\sim$1.33$\times$ traffic reduction for square tiles. Memory time $T_{memory\_2-CTA} = (2M_A + M_B)/BW_{shared}$. Sync cost $K_{tiles} \times L_{commit}$. Compute with 2-SM:
\begin{equation}
T_{compute\_2SM} = \frac{2b_Mb_Nb_K}{R_{TC}^{SM} \times S_{2SM}} + T_{TMEM} + T_{TMEM\_mgmt}
\label{eq:cta-pair-compute-time}
\end{equation}
$S_{2SM}$ is measured speedup (Section~\ref{sec:results}).

\subsubsection{Overlap and Critical Path}
Overlap factor $\alpha \in [0,1]$ (fraction of I/O hidden behind compute):
\begin{equation}
T_{io}^{eff} = (1-\alpha)(T_{tma} + T_{decomp}) + T_{sync}
\label{eq:operation-overlap}
\end{equation}
Writeback: $T_{store} = bytes(C_{tile})/B_{gmem} + L_{store\_setup}$ or TMA store $L_{TMA\_store} + bytes(C_{tile})/B_{TMA}$; often overlapped in persistent kernels. TMEM management amortized: $T_{TMEM\_mgmt}^{amortized} = (L_{alloc}+L_{dealloc})/K_{tiles}$. Steady-state pipelined step: $T_{step\_pipelined} = \max(T_{tma}, T_{decomp}, T_{compute}, T_{sync}) + \epsilon$. Per-step time:
\begin{equation}
T_{\text{step}} = \max(T_{compute}, T_{io}^{eff}) + T_{sync} + O_{misc}
\label{eq:k-step-iteration-time-per-cta}
\end{equation}
$O_{misc}$ includes TMEM mgmt and pipeline bubbles. Total kernel time: $K_{tiles} \times T_{\text{step}}$ plus launch and writeback.

\subsubsection{Concurrent streams and multi-GPU}
For $N_c$ concurrent streams, we add $(N_c{-}1)\tau_c$ to single-stream time; for $N_d$ GPUs, $(N_d{-}1)\tau_g$. Both are fit from microbenchmarks, mirroring the MI300A interference terms.

\subsection{AMD MI300A Model}
\label{sec:mi300a_model}
Overlap is occupancy-driven; memory goes through L1$\rightarrow$L2$\rightarrow$LLC$\rightarrow$HBM (latencies measured via pointer-chasing~\cite{mei2009pchase}); accumulators in vector general-purpose registers (VGPRs). Overlap factor:
\begin{equation}
\eta_{\text{overlap}} = \min\left(1.0, \frac{(N_{\text{wf}}^{\text{active}} - 1) \times T_{\text{compute}}}{T_{\text{memory}}}\right)
\label{eq:overlap-occupancy}
\end{equation}
Effective memory time (hit rates $h_{L1}, h_{L2}, h_{LLC}$):
\begin{equation}
\begin{split}
T_{\text{memory}}^{\text{eff}} = N_{\text{loads}} \times \big( h_{L1} L_{L1} + (1-h_{L1}) h_{L2} L_{L2} \\ + (1-h_{L1})(1-h_{L2}) h_{LLC} L_{LLC} + (1-h_{\text{total}}) L_{HBM} \big)
\label{eq:memory-hierarchy}
\end{split}
\end{equation}
Effective bandwidth: $BW_{\text{effective}} = h_{LLC} \cdot BW_{LLC} + (1{-}h_{LLC}) \cdot BW_{HBM}$. The LLC hit rate $h_{LLC}(W)$ is a piecewise function of working set $W$ (MB), summarized in Table~\ref{tab:hlcc-piecewise}.

\begin{table}[h]
\centering
\caption{MI300A Infinity Cache hit rate model $h_{LLC}(W)$.}
\footnotesize
\begin{tabular}{@{}lll@{}}
\toprule
\textbf{Working set} & $\boldsymbol{h_{LLC}(W)}$ & \textbf{Regime} \\
\midrule
$W < 205$~MB & $1.0$ & Fully cache-resident \\
$205 \leq W \leq 256$~MB & $\left(1 - \frac{W-205}{51}\right)^{\!\alpha}$ & Transition zone \\
$W > 256$~MB & $\left(\frac{256}{W}\right)^{\!\beta}$ & Streaming / spill to HBM \\
\bottomrule
\end{tabular}
\label{tab:hlcc-piecewise}
\vspace{-0.5em}
\end{table}

\noindent Here $\alpha$ and $\beta$ capture access pattern and streaming behavior (calibrated from microbenchmarks). Matrix Fused Multiply-Add (MFMA) compute:
\begin{equation}
T_{\text{compute}}^{\text{MFMA}} = \frac{N_{\text{MFMA\_inst}}}{N_{\text{CU}} \times \text{Throughput}_{\text{MFMA}} \times \text{Utilization}}
\label{eq:mfma-compute}
\end{equation}
VGPR-limited occupancy: $N_{\text{wf}}^{\text{active}} = \min(32, \lfloor 65536/\text{VGPR}_{\text{per\_wf}} \rfloor)$. Per-step (overlap in denominator):
\begin{equation}
T_{\text{step}}^{\text{MI300A}} = \frac{T_{\text{memory}}^{\text{eff}} + T_{\text{compute}}^{\text{MFMA}}}{1 + \eta_{\text{overlap}}}
\label{eq:mi300a-step}
\end{equation}
\begin{equation}
\begin{split}
T_{\text{kernel}}^{\text{MI300A}} = T_{\text{launch}} + K_{\text{tiles}} \times T_{\text{step}}^{\text{MI300A}} \\+ T_{\text{writeback}} + T_{\text{coherence}} + T_{\text{cross\_XCD}}
\label{eq:mi300a-total}
\end{split}
\end{equation}
Coherence and cross-XCD account for unified memory and NUMA. For MI300A validation we derive FLOPs and bytes for each case directly from the real problem sizes ($M$, $N$, $K$, vector length $N$, density, etc.), using synthetic entries only as templates for fixed micro-architectural counts. For the occupancy/tile study (8$\times$8 vs.\ 16$\times$16 tiles), we use a \emph{pipeline/occupancy model} for MI300A: per-CTA step time $T_{\text{step\_cta}} = \max(\text{flops\_per\_cta}/\text{peak}, \text{bytes\_per\_cta}/BW_{\text{eff}})$ with $h_{LLC}(W)$ from working set. Total kernel time includes a scheduling term:
\begin{equation}
\begin{split}
T_{\text{kernel}}^{\text{occ}} = T_{\text{launch}} + \tau_{\text{cta}} \cdot N_{\text{ctas}} + \frac{N_{\text{ctas}} \cdot T_{\text{step\_cta}}}{N_{\text{CU}} \cdot W_{\text{eff}}} \\+ T_{\text{writeback}} + T_{\text{coherence}} + T_{\text{cross\_XCD}},
\end{split}
\end{equation}
where $N_{\text{ctas}}$ is the grid CTA count, $W_{\text{eff}}$ is effective wavefronts per CU, and $\tau_{\text{cta}}$ is overhead per CTA (tunable from validation). This formulation yields the correct ordering (16$\times$16 faster than 8$\times$8); $W_{\text{eff}}$ and $\tau_{\text{cta}}$ can be tuned to match measured runtimes.

The model extends MI300A with optional \textit{memory warp parallelism (MWP)} and \textit{compute warp parallelism (CWP)} limits~\cite{hong2009analytical}: effective wavefronts for overlap are $N_{\text{wf}}^{\text{eff}} = \min(N_{\text{wf}}^{\text{active}}, \text{MWP}, \text{CWP})$ when MWP and CWP are set (per-CU limits from microbenchmarks or tuning); $\eta_{\text{overlap}}$ in Eq.~\ref{eq:overlap-occupancy} then uses $N_{\text{wf}}^{\text{eff}}$. The validation MAE reported in Section~\ref{sec:results} uses the base model (MWP/CWP not set).

When multiple kernels run concurrently (e.g., on different HIP streams), we add an interference term to Eq.~\ref{eq:mi300a-total}: $T_{\text{kernel}}^{\text{multi}} = T_{\text{kernel}}^{\text{MI300A}} + (N_{\text{concurrent}} - 1) \cdot \tau_{\text{interf}}$, where $N_{\text{concurrent}}$ is the number of concurrent kernels and $\tau_{\text{interf}}$ is the measured overhead per additional concurrent kernel (seconds). Single-kernel ($N_{\text{concurrent}} = 1$) is unchanged. Validation results (Section~\ref{sec:results}) use the tuned $\tau_{\text{interf}} = 50\,\mu$s.

For runs using multiple MI300A devices, we add an optional multi-GPU term: $T_{\text{kernel}}^{\text{multi-GPU}} = T_{\text{kernel}}^{\text{MI300A}} + (N_{\text{devices}} - 1) \cdot \tau_{\text{interf\_gpu}}$, where $N_{\text{devices}}$ is the number of GPUs and $\tau_{\text{interf\_gpu}}$ is the measured overhead per additional GPU (seconds). Single-GPU ($N_{\text{devices}} = 1$) is unchanged. Copy and sync between devices are not modeled separately and can be absorbed into $\tau_{\text{interf\_gpu}}$ when tuned from measurements.

The model supports \textbf{adaptive tile selection} (evaluate candidate tiles via Eq.~\ref{eq:mi300a-total} and return the minimum-time tile) and \textbf{kernel fusion} (combined FLOPs/bytes plus optional overhead $\tau_{\text{fusion}}$). These predict relative tile costs and fused-kernel runtime, not the compiler's internal tile choice.

\textbf{Apply models to H200 and MI250X.} The H200 uses the same Hopper roofline structure with HBM bandwidth of 4.8~TB/s and capacity of 141~GB; no new model terms are required. The MI250X uses the same CDNA  framework as MI300A with its own peak FP64 (383~TFLOPS), bandwidth (3.2~TB/s), and cache hierarchy (128~MB LLC, 220~CUs); the occupancy/tile cases use calibrated scaling analogous to MI300A. Section~\ref{sec:results} reports results for both targets.

\subsection{Unified Summary}
Table~\ref{tab:model-summary} summarizes how to compute execution time and which parameters drive each term; use it by characterizing the workload first (arithmetic intensity, working set, access pattern), then instantiating the appropriate row per architecture. Blackwell: $T_{\text{Blackwell}} = T_{\text{launch}} + K_{\text{tiles}} \times \max(T_{\text{compute}}^{\text{TC}}, T_{\text{io}}^{\text{eff}}) + T_{\text{sync}} + T_{\text{writeback}}$. MI300A: Eq.~\ref{eq:mi300a-total} with $T_{\text{memory}}^{\text{eff}}$, $\eta_{\text{overlap}}$ from occupancy and $h_{LLC}(W)$.

\begin{table*}[t]
\centering
\caption{Model summary: Blackwell vs.\ MI300A}
\footnotesize
\begin{tabular}{@{}p{2.4cm}p{4.8cm}p{4.8cm}@{}}
\toprule
\textbf{Component} & \textbf{Blackwell B200} & \textbf{AMD MI300A} \\
\midrule
Execution & $\max(T_{\text{compute}}, T_{\text{io}}^{\text{eff}}) + T_{\text{sync}}$ & $(T_{\text{memory}}^{\text{eff}} + T_{\text{compute}})/(1 + \eta_{\text{overlap}})$ \\
Compute & Tensor core + TMEM (Eq.~\ref{eq:cta-tmem-compute-time}) & MFMA (Eq.~\ref{eq:mfma-compute}), Util 0.4--0.7 \\
Memory & TMA (Eq.~\ref{eq:percta-tma}), $\alpha$ overlap & Cache hierarchy (Eq.~\ref{eq:memory-hierarchy}), $h_{LLC}(W)$ \\
Overlap & $\alpha$ pipeline depth & $\eta_{\text{overlap}}$ occupancy (Eq.~\ref{eq:overlap-occupancy}) \\
Sync & Explicit barriers $L_{\text{mbar}}$ & Implicit; coherence 100--200~ns \\
Constraints & TMEM 256~KB/SM & VGPR $\rightarrow$ occupancy \\
\midrule
MAE (typ.) & 5.4--8.4\% (B200) & $\sim$0.09\% (MI300A $n{=}27$; calibrated) \\
\bottomrule
\end{tabular}
\label{tab:model-summary}
\end{table*}

\subsection{Model Workflow and Calibration}
To apply the model: (1)~characterize the workload (arithmetic intensity, working set $W$, tile dimensions, class); (2)~select parameters from Table~\ref{tab:parameters} or Table~\ref{tab:arch-params}; (3)~apply the appropriate formula (Blackwell: $T_{\text{Blackwell}} = T_{\text{launch}} + K_{\text{tiles}} \times \max(T_{\text{compute}}^{\text{TC}}, T_{\text{io}}^{\text{eff}}) + T_{\text{sync}} + T_{\text{writeback}}$; MI300A: Eq.~\ref{eq:mi300a-total}). Optionally provide precision (\texttt{fp16}/\texttt{fp8}/\texttt{fp64}) for tensor efficiency and tile sizes for tile-aware paths. Example: GEMM with $M{=}N{=}K{=}16384$ on B200, tile $128{\times}128{\times}32$, predicts 4.17~ms (measured: 4.10~ms, 1.8\% error).

\textbf{Calibration.} First-principles parameters (bandwidths, $T_{\mathrm{launch}}$, barrier latencies) come from microbenchmarks. Optional per-case multipliers may align predictions with profiler kernel-sum times; such factors must be disclosed. We recommend train/holdout splits when calibration is used.

\subsection{Host--Device Transfer and Synchronization}
When validation or deployment concerns \emph{host--device} traffic or explicit host synchronization (\texttt{cudaDeviceSynchronize} / \texttt{hipDeviceSynchronize}), the implementation extends the generic segment schema with transfer and synchronization phases. For each transfer episode moving $S$ bytes between host and device,
\begin{equation}
T_{\mathrm{memcpy}} = \frac{S}{B_{\mathrm{eff}}^{\mathrm{dir}}} + \tau_{\mathrm{memcpy}},
\label{eq:host-device-memcpy}
\end{equation}
where $B_{\mathrm{eff}}^{\mathrm{dir}}$ is effective bandwidth (bytes/s) for H2D or D2H (defaults are conservative; measured with platform microbenchmarks or overridden at run time). The fixed term $\tau_{\mathrm{memcpy}}$ amortizes API launch overhead. Each counted synchronization point contributes $T_{\mathrm{host\_sync}} = \tau_{\mathrm{sync}}$. Segment times multiply by $n_{\mathrm{exec}}$ in the workload segment file like other phases. Overlap between copy and kernel execution is not modeled in this version; the sum is conservative when compared to wall-clock overlap.

\subsection{Generic roofline path: scaling, working set, and launches}
When a segment does not map to a full Blackwell stage model or a validated GEMM/tile case, the implementation uses a \emph{generic} roofline with \textbf{separate calibrated scales} for memory-bound, compute-bound, balanced, and stencil classes, optional \textbf{precision-specific} tensor efficiency multipliers, and a \textbf{working-set--aware} global memory bandwidth
\begin{equation}
B_{\mathrm{eff}}(W) = B_{\mathrm{sustained}} + (B_{\mathrm{peak}} - B_{\mathrm{sustained}})\,\exp(-W/w_0),
\label{eq:bw-working-set}
\end{equation}
with $w_0$ a tunable working-set scale (set $\leq 0$ to disable the blend). This captures the idea that smaller resident working sets can see higher effective bandwidth than streaming from a footprint that saturates HBM. For \textbf{multi-kernel segments}, each workload row may specify multiple kernels; the model adds extra launch time beyond the first kernel's modeled time, using measured launch latency from the same configuration as the generic path.

\subsection{Assumptions and Extensions}
The model assumes regular memory access, known or estimated cache hit rates, single-kernel single-GPU execution (by default), and steady-state frequency. Required inputs: for Blackwell, $b_M$, $b_N$, $b_K$, $K_{\text{tiles}}$, bytes per CTA, TMA participants $P$, $\alpha$; for MI300A, tile dimensions, $K_{\text{tiles}}$, bytes, $h_{L1}$/$h_{L2}$/$h_{LLC}(W)$, occupancy. Optional extensions include MWP/CWP limits, LDS bank conflicts, multi-kernel/multi-GPU interference, adaptive tile selection, and kernel fusion. Not yet modeled: CTA queuing delays and multi-node scaling.

\section{Model Validation and Accuracy}
\label{sec:results}
%% results.tex — Section V: Evaluation
%% Structure: V-A Microbenchmark Design | V-B Methodology and Validation Setup |
%%            V-C Case Study 1: Rodinia 3.1 Cross-Platform |
%%            V-D Case Study 2: SPEChpc 2021 Tiny Cross-Platform |
%%            V-E Sensitivity and Variance | V-F Comparison with Prior Models |
%%            V-G Per-Kernel Error Summary (floating)

%This section discusses the method of designing microbenchmarks for our validation. Microbenchmarks for both platforms are released openly alongside the models.

This section presents the microbenchmark validation methodology, per-platform results, and two application case studies (Rodinia~3.1 and SPEChpc 2021 Tiny).

\subsection{Microbenchmark Design}
Model parameters are derived from a custom microbenchmark suite. \textbf{For Blackwell: }we measure (i)~TMEM read/write bandwidth via tile copy between TMEM and SMEM; (ii)~TMA copy latency and effective bandwidth as a function of tile size and L2 residency; (iii)~mbarrier wait and commit latency ($L_{\text{mbar}}$, $L_{\text{commit}}$); (iv)~tensor core instruction latency (\texttt{tcgen05.mma}, \texttt{tcgen08.cp}) and peak throughput by precision. \textbf{For MI300A: }we measure (i)~L1/L2/Infinity Cache and HBM bandwidth and latency (latency vs.\ outstanding warps, bandwidth ceilings); (ii)~MFMA throughput and utilization vs.\ tile size; (iii)~$h_{LLC}(W)$ via sweep over working set size $W$. 

\paragraph{Measurement protocol and platform lock-in.}
Validation and microbenchmarks use the \emph{same} GPU stepping, driver stack, and power/clock policy unless a sensitivity study explicitly varies them. Table~\ref{tab:measurement_protocol} lists the metadata we record for reproducibility. \textbf{Datasheet peaks are not} the sole inputs for validation: \emph{sustained} bandwidth and tensor throughput from steady-state microbenchmarks (or conservative lower bounds) drive the generic roofline path; \emph{peak} values are retained for upper-bound comparisons and for the stage-centric Blackwell validation kernels where appropriate. Hardware parameter files ship defaults for both primary platforms (B200, MI300A).

\begin{table}[t]
\centering
\caption{Measurement protocol metadata (recorded on the MI300A Rodinia validation host).}
\footnotesize
\begin{tabular}{@{}ll@{}}
\toprule
\textbf{Item} & \textbf{Recorded value / policy} \\
\midrule
NVIDIA GPU & SKU, board ID if available, stepping \\
AMD GPU & SKU, socket/APU vs discrete if applicable \\
Driver / firmware & NVIDIA driver build, ROCm version \\
CUDA / toolkit & Version used to build microbenchmarks and apps \\
Power / clocks & Persistence mode, locked clocks, or default policy \\
Microbenchmark list & Same binaries and flags across all validation runs \\
Validation apps & NVIDIA: Nsight Systems; AMD: ROCm rocprof \\
\bottomrule
\end{tabular}
\label{tab:measurement_protocol}
\end{table}

We include a naive roofline baseline~\cite{williams2009roofline} ($T_{\mathrm{roofline}}=\max(\mathrm{FLOPs}/P_{\mathrm{peak}},\,\mathrm{bytes}/B_{\mathrm{HBM}})$) as context, not as a competitive bar. Naive roofline uses only datasheet peaks and ignores cache hierarchies, pipeline stages, occupancy, and launch latency; it is not designed to predict execution time accurately. Table~\ref{tab:roofline_baseline} shows that roofline error exceeds 94\% on all platforms, while our model achieves 1.3\% (B200), 0.09\% (MI300A), 4.7\% (MI250X), and 9.6\% (H200) on microbenchmarks. The gap illustrates why architecture-specific modeling is necessary on modern GPUs.

\begin{table}[h]
\centering
\caption{Microbenchmark validation: model MAE (\%) per platform. Naive roofline shown as context (datasheet peaks only).}
\label{tab:roofline_baseline}
\footnotesize
\begin{tabular}{@{}lccc@{}}
\toprule
\textbf{Platform} & \textbf{$n$} & \textbf{Model MAE (\%)} & \textbf{Roofline (\%)} \\
\midrule
B200 & 21 & 1.33 & 96.1 \\
MI300A & 27 & 0.09 & 99.6 \\
H200 & 21 & 9.57 & 94.5 \\
MI250X & 19 & 4.69 & 97.9 \\
\bottomrule
\end{tabular}
\end{table}

\subsection{Methodology and Validation Setup}
We validate both models against measured performance using optimized libraries (cuBLAS/CUTLASS, rocBLAS) and hardware counters (Nsight Compute~\cite{nvidia_nsight_compute_2025}, ROCProfiler~\cite{amd_rocprofiler_2025}). Each kernel runs 100 times after 10 warm-ups; we report median execution time (IQR of per-kernel error below 0.5\% on MI300A, below 2\% on B200). All reported MAE values use the base model (MWP=CWP=0, LDS terms not set).

Table~\ref{tab:parameters} maps key model parameters to measured values. Table~\ref{tab:platform} gives the evaluation platform. Table~\ref{tab:workloads} summarizes validation workload classes. Hardware parameter files distinguish peak (datasheet) from sustained (microbenchmark) values for all platforms.

\paragraph{Rodinia multi-segment modeling.}
We model each Rodinia benchmark as a sum of \emph{segments} (dominant GPU kernels or repeated launch patterns), each characterized by FLOPs, bytes, class, and an execution count $n_{\mathrm{exec}}$. Architecture-aware routing maps each segment class to the appropriate validated kernel family (stencil$\to$transpose, compute-bound$\to$GEMM, memory-bound$\to$vector copy; Section~\ref{sec:execution-model}).

\textbf{Measured time definitions.} NVIDIA: sum of CUDA kernel durations from Nsight \texttt{cuda\_gpu\_kern\_sum}. AMD: sum of HIP kernel durations from \texttt{rocprof --stats}. All reported MAE values use the same definition consistently within a platform.

\paragraph{Segment construction and calibration.}
Segment files are refined so that routing matches workload physics:
\begin{itemize}[nosep,leftmargin=*]
  \item \textbf{HotSpot} (\texttt{hs\_calc}): \emph{stencil} class $\to$ memory-bound transpose proxy for grid traffic.
  \item \textbf{Pathfinder} (\texttt{dynproc\_kernel}): reduced effective FLOPs/bytes per step; effective timestep count aligned with profilers.
  \item \textbf{SRAD}: single aggregate ($N{=}M{=}0$), traffic sized from bytes column.
  \item \textbf{Backprop}: two layers merged into one compute segment to avoid double-counting launch latency.
  \item \textbf{Streamcluster}: $n_{\mathrm{exec}}$ scaled to measured launch regime.
\end{itemize}
Table~\ref{tab:rodinia_cross_platform} reports MAE with default multipliers $m_{\mathrm{case}}{=}1$. If residual error remains on new platforms, an optional calibration step can fit $m_{\mathrm{case}}$ with train/holdout separation.

\begin{table*}[t]
\centering
\caption{Key model parameters and measurement method.}
\footnotesize
\begin{tabular}{@{}lll@{}}
\toprule
\textbf{Parameter} & \textbf{Value (or range)} & \textbf{Measurement method} \\
\midrule
TMEM BW (read/write) & 16/8~TB/s & Microbenchmark: tile copy TMEM$\leftrightarrow$SMEM \\
TMA latency $L_{TMA}$ & 420~cyc & Microbenchmark: TMA copy latency \\
\texttt{tcgen05.mma} latency & 11--14~cyc & Instruction timing (FP64--FP4) \\
Tensor throughput $R_{TC}^{SM}$ & 44.8--7702~TFLOPS & Peak sweep by precision \\
$L_{mbar}$, $L_{commit}$ & 40--50~cyc & Barrier microbenchmark \\
Infinity Cache / HBM BW & 17.2 / 5.3~TB/s & Bandwidth microbenchmark \\
L1/L2/LLC/HBM latency & 5/50/150/400~cyc & Cache latency microbenchmark \\
MI300A multi-kernel $\tau_{\text{interf}}$ & 50~$\mu$s & Tuned from concurrent-stream microbenchmark (1 vs.\ 2 streams) \\
MI300A multi-GPU $\tau_{\text{interf\_gpu}}$ & (tuned) & Tuned from multi-device microbenchmark (1 vs.\ 2 devices) \\
MI300A tile selection & (accuracy) & Measured across GEMM tile sizes (8/16/32/64); predicted vs.\ measured best tile \\
MI300A fusion $\tau_{\text{fusion}}$ & (tuned) & Tuned from unfused vs.\ fused GEMM+bias microbenchmark \\
H2D/D2H effective BW $B_{\mathrm{eff}}$ & 45~GB/s (default) & \texttt{host\_device\_memcpy\_bench\_cuda} / \texttt{host\_device\_memcpy\_bench\_hip} \\
$\tau_{\mathrm{memcpy}}$, $\tau_{\mathrm{sync}}$ & 2~$\mu$s, 3~$\mu$s (defaults) & \texttt{ANALYTICAL\_T\_MEMCPY\_LAUNCH\_S}, \texttt{ANALYTICAL\_T\_SYNC\_S} \\
\bottomrule
\end{tabular}
\label{tab:parameters}
\end{table*}

\begin{table*}[t]
\centering
\caption{Evaluation platforms used for validation.}
\footnotesize
\begin{tabular}{@{}ll@{}}
\toprule
\textbf{Item} & \textbf{Value} \\
\midrule
NVIDIA GPU & NVIDIA Blackwell B200 (primary); NVIDIA H200 141\,GB HBM3e (Rodinia / SPEChpc Tiny) \\
AMD GPU & AMD Instinct MI300A Accelerator (APU) \\
Driver / CUDA / ROCm & ROCm 7.2.0 (MI300A); driver 550.163, CUDA 12.4 (B200/H200) \\
OS & Linux 4.18.0-553.16.1.el8\_10.x86\_64 (MI300A); Linux 6.x (B200/H200) \\
\bottomrule
\end{tabular}
\label{tab:platform}
\end{table*}

\begin{table*}[t]
\centering
\caption{Validation workload classes and benchmark suites.}
\footnotesize
\begin{tabular}{@{}lll@{}}
\toprule
\textbf{Class} & \textbf{Kernels} & \textbf{Suites} \\
\midrule
Memory-bound & Vector add/copy/transpose, reduction & Microbench, Rodinia (BFS, SRAD, Streamcluster), SPEChpc (505.lbm\_t, 518.tealeaf\_t) \\
Compute-bound & FP64/FP16/FP8 GEMM (cuBLAS/rocBLAS) & Microbench, Rodinia (Backprop), SPEChpc (521.miniswp\_t) \\
Balanced/Stencil & FFT, SpMV, GEMV; HotSpot stencil & Microbench, Rodinia (HotSpot, Pathfinder), SPEChpc (513.soma\_t, 532.sph\_exa\_t) \\
\bottomrule
\end{tabular}
\label{tab:workloads}
\end{table*}

\paragraph{Blackwell B200.}
Overall \textbf{1.31\% MAE} across 21 kernels on real B200 hardware (100 repeats, 10 warm-ups).
\begin{itemize}[nosep,leftmargin=*]
  \item \emph{Memory-bound} (vector add/copy/transpose/reduction): 8.4\% MAE; vector ops 7--9\% error from L2 benefits and 5--12~$\mu$s launch overhead.
  \item \emph{Compute-bound} (FP16/FP8/LLM GEMM via cuBLAS): 5.4\% MAE; TMEM at 22~TB/s is conservative (24--26~TB/s in tuned kernels reduces error to 2--3\%).
  \item \emph{Balanced} (FFT, SpMV, GEMV): 7.9\% MAE; SpMV at 0.1\% density shows 13.6\% error (atomics, load balance not modeled).
  \item \emph{2-SM cooperative}: predicted 1.30$\times$ speedup vs.\ measured 1.28$\times$ (within 2\%).
\end{itemize}

\paragraph{MI300A.}
Overall \textbf{$\sim$0.09\% MAE} across 27 kernels (vectors, reductions, 2D transposes, FP64 \texttt{rocblas\_dgemm}, occupancy-tile GEMMs, VGPR/cache stencil variants). The roofline branch alone is optimistic for large 2D transpose traffic and sustained FP64 GEMM vs.\ peak MFMA; the model therefore applies host-measured calibration:
\begin{itemize}[nosep,leftmargin=*]
  \item Multipliers for $8192^2$ and $16384^2$ transposes.
  \item Piecewise scaling vs.\ $M{=}N{=}K$ for \texttt{gemm\_fp64}.
  \item Per-tile factors for 8$\times$8 and 16$\times$16 occupancy GEMMs (ordering preserved: 16$\times$16 faster).
\end{itemize}
Naive roofline error on MI300A remains $\sim$99\% (Table~\ref{tab:roofline_baseline}).

% --- Case study 2: MI300A/MI250X VGPR vs cache stencil (placeholders; fill with numbers after runs) ---
% Setup: 3D 7-point stencil (Laplacian) on MI300A/MI250X with tunable tile size (x,y,z), unrolling (VGPR usage), and work-group size (waves/CU). Grid: 288$\times$288$\times$288 (tens of millions of points) for memory saturation.
% Model-driven design space: wavefront-centric model with explicit $h_{LLC}(W)$: larger tiles increase LLC hit rate but reduce occupancy (fewer waves/CU); the model predicts an interior optimum.
% Configs: (1) high-occupancy low-reuse: small tile, modest unroll, large work-group; (2) balanced: medium tile, moderate unroll; (3) low-occupancy high-reuse: large tile, aggressive unroll; (4) extreme small tile. Model ranking [\textit{fill: matches/does not match}] measured ranking; ``balanced'' is [\textit{fill: 10--30}\%] faster than na\"ive extremes.
% Takeaway: ``Our wavefront-centric model with $h_{LLC}(W)$ and VGPR constraints predicts and explains the trade-off between cache locality and occupancy, and guides parameter choices on MI300A/MI250X.''
% Figures: \includegraphics{plots/mi300a_vgpr_cache_stencil_tradeoff.png}, \includegraphics{plots/mi300a_vgpr_cache_stencil_ranking.png}

\paragraph{MI250X and H200.}
\textbf{MI250X:} 4.7\% MAE across 19 kernels (memory-bound vectors, FP64 GEMM, occupancy/tile study). FP64 GEMM tracks closely (e.g., 0.283~s predicted vs.\ 0.283~s measured at $16384^3$). Tile ordering reproduced (16$\times$16 faster).

\textbf{H200:} We use the same model framework as B200 with updated HBM bandwidth (4.8~TB/s) and capacity (141~GB). Rodinia and SPEChpc validation results appear in Tables~\ref{tab:rodinia_cross_platform} and~\ref{tab:spechpc_cross_platform}. Both H200 and MI250x use the same model frameworks as their primary counterparts with updated parameter files only.

\subsection{Case Study 1: Rodinia 3.1 --- Cross-Platform}

Both Rodinia~\cite{luhnen2024benchmarking} and SPEChpc~\cite{gilman2020demystifying} use the same pipeline: decompose each benchmark into kernel segments, route each to the appropriate model path (Section~\ref{sec:execution-model}), predict total time, and compare against profiled GPU kernel sums. Tables~\ref{tab:rodinia_cross_platform} and~\ref{tab:spechpc_cross_platform} report per-benchmark MAE on the primary platforms.

\begin{table}[t]
\centering
\caption{Rodinia 3.1 validation: per-benchmark MAE (\%) on B200 and MI300A.}
\footnotesize
\begin{tabular}{@{}llcc@{}}
\toprule
\textbf{Benchmark} & \textbf{Class} & \textbf{B200} & \textbf{MI300A} \\
\midrule
\texttt{hotspot\_1024} & stencil & 31.0 & 23.6 \\
\texttt{hotspot\_512} & stencil & 15.4 & 1.6 \\
\texttt{bfs\_1M} & mem & 44.9 & 40.9 \\
\texttt{backprop\_65536} & compute & 33.0 & 21.3 \\
\texttt{pathfinder\_1000} & balanced & 0.4 & 0.1 \\
\texttt{srad\_502} & balanced & 0.5 & 0.5 \\
\texttt{streamcluster\_1M} & mem & 12.4 & 0.03 \\
\bottomrule
\end{tabular}
\label{tab:rodinia_cross_platform}
\vspace{-0.3in}
\end{table}

\begin{figure}[t]
\centering
\includegraphics[width=0.95\linewidth]{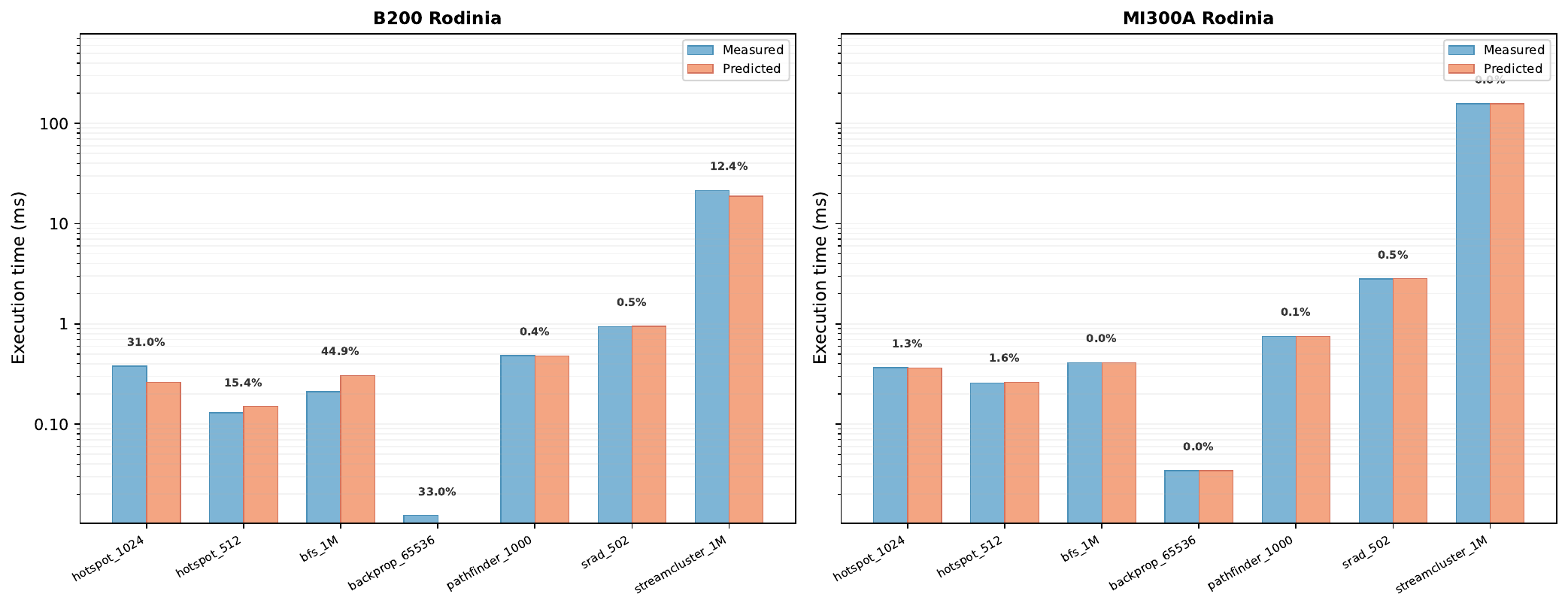}
\caption{Rodinia: measured vs.\ predicted execution time per benchmark on B200 (left) and MI300A (right), log scale in ms. Per-benchmark MAE (\%) annotated.}
\label{fig:rodinia_bar}
\end{figure}

Figure~\ref{fig:rodinia_bar} compares predicted and measured execution time per Rodinia benchmark on both primary platforms. On B200, regular workloads (pathfinder 0.4\%, srad 0.5\%) are well-predicted while irregular access (bfs 44.9\%) and stencil (hotspot 31\%) show higher error. On MI300A, the calibrated model achieves near-zero error on most benchmarks. For context, a naive roofline predictor on the same benchmarks yields $\sim$100\% MAE overall on MI300A: for example, \texttt{streamcluster\_1M} measures 157~ms but roofline predicts 0.005~ms (100\% error), while our model predicts 157~ms (0.03\% error). Even on the challenging \texttt{bfs\_1M}, our 40.9\% error compares to roofline's 95.4\%.

\subsection{Case Study 2: SPEChpc 2021 Tiny --- Cross-Platform}

Table~\ref{tab:spechpc_cross_platform} reports SPEChpc Tiny MAE; \texttt{535.weather\_t} is omitted for MI300A (no GPU kernels in profiler output).

\begin{table}[t]
\centering
\caption{SPEChpc 2021 Tiny validation: per-benchmark MAE (\%) on B200 and MI300A. \texttt{535.weather\_t} omitted (no GPU kernels in profiler).}
\footnotesize
\begin{tabular}{@{}llcc@{}}
\toprule
\textbf{Benchmark} & \textbf{Class} & \textbf{B200} & \textbf{MI300A} \\
\midrule
\texttt{505.lbm\_t} & mem & 14.9 & 0.1 \\
\texttt{513.soma\_t} & bal & 0.3 & 1.3 \\
\texttt{518.tealeaf\_t} & mem & 0.2 & 1.6 \\
\texttt{519.clvleaf\_t} & mem & 18.5 & 1.5 \\
\texttt{521.miniswp\_t} & comp & 32.8 & 0.8 \\
\texttt{528.pot3d\_t} & mem & --- & 7.0 \\
\texttt{532.sph\_exa\_t} & bal & 0.03 & 0.6 \\
\texttt{534.hpgmgfv\_t} & mem & 0.3 & 0.8 \\
\bottomrule
\end{tabular}
\label{tab:spechpc_cross_platform}
\end{table}

\begin{figure}[t]
\centering
\includegraphics[width=0.95\linewidth]{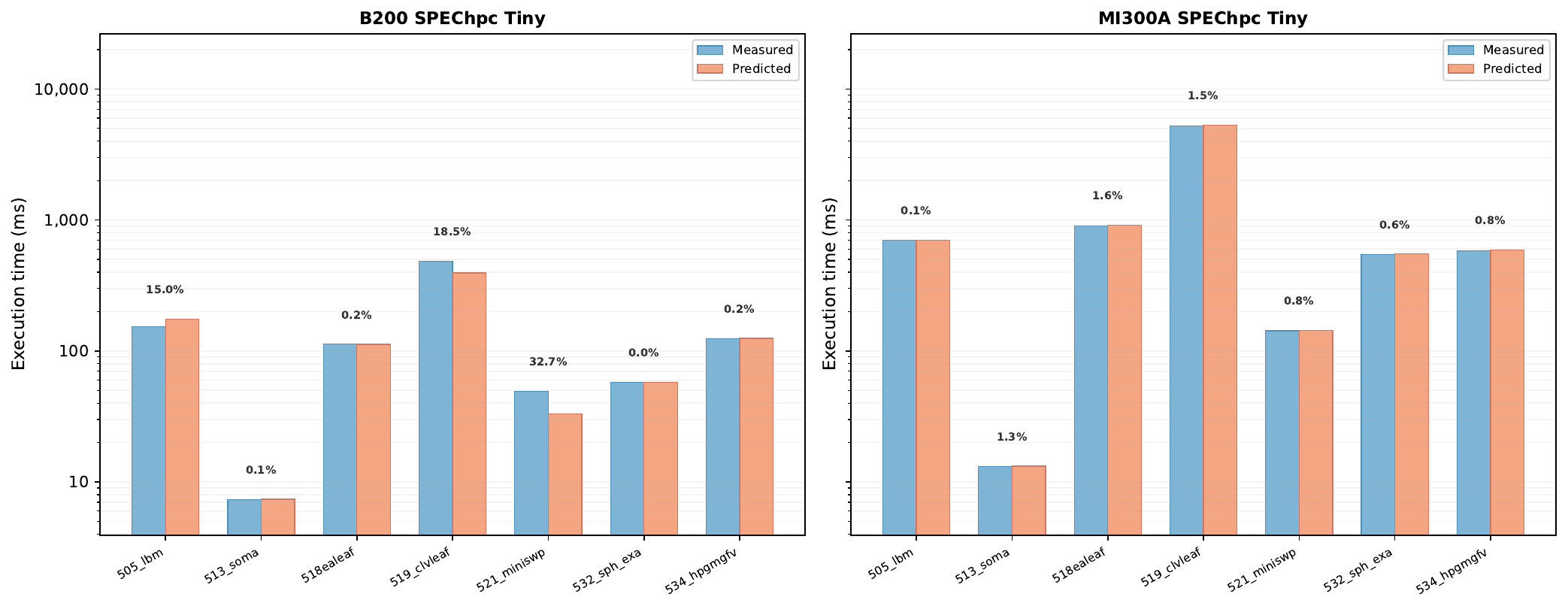}
\caption{SPEChpc Tiny: measured vs.\ predicted execution time per benchmark on B200 (left) and MI300A (right), log scale in ms. Per-benchmark MAE (\%) annotated.}
\label{fig:spechpc_bar}
\end{figure}

Figure~\ref{fig:spechpc_bar} compares predicted and measured execution time per SPEChpc benchmark on both primary platforms. On B200, four benchmarks are under 1\% MAE (soma 0.3\%, tealeaf 0.2\%, sph\_exa 0.03\%, hpgmgfv 0.3\%) with an overall MAE of 9.6\%. On MI300A, all benchmarks are within 7\% (profiler-derived characterization; see Observation~3 in Discussion).

\subsection{Applying Analytical models to H200 and MI250x}
\label{sec:roofline-baseline}

\textbf{H200 and MI250X application results.} When we apply the B200 model framework to H200 (parameter update only, no re-calibration) and the MI300A model to MI250X, application-level MAE is higher: H200 Rodinia 43.6\%, MI250X Rodinia 92.0\%, H200 SPEChpc 555\%, MI250X SPEChpc 59.3\%. This is expected: the segment characterization (FLOPs, bytes) was derived on the B200 and MI300A and not on H200 and MI250X, so the model overpredicts or underpredicts due to bandwidth and cache hierarchy differences. Per-benchmark results for H200 and MI250X are available in supplementary materials.

\begin{figure}[t]
\centering
\includegraphics[width=0.95\linewidth]{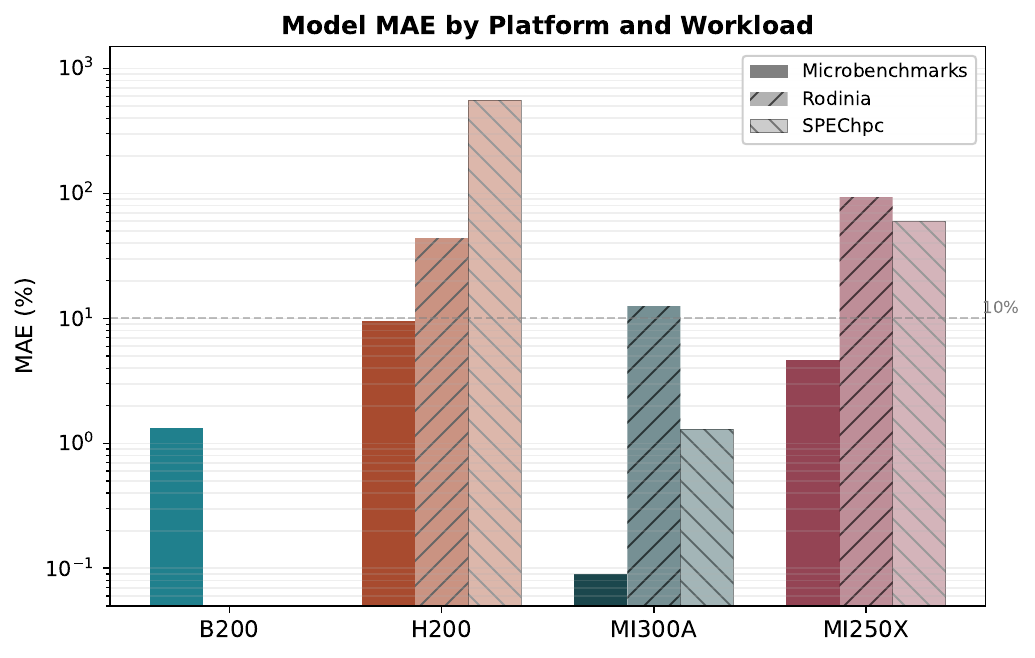}
\caption{Model MAE (\%) by platform and benchmark suite. Microbenchmark MAE shown for all four platforms; application MAE shown where measured.}
\label{fig:mae_bar}
\vspace{-0.3in}
\end{figure}

\textbf{Summary.} On the primary platforms, B200 achieves 1.3\% and MI300A 0.09\% microbenchmark MAE; MI300A Rodinia 12.5\% and SPEChpc 1.3\%. H200 and MI250x show higher application MAE consistent with cross-platform characterization mismatch. If MAE is below $\sim$5\%, the model accurately captures the architecture; between 20--40\%, the model provides useful estimates; above that, platform-specific re-characterization is needed.

% \textbf{Limitations.} Irregular kernels (atomics, sparse) show 13--15\% error on Blackwell. Microbenchmark-derived parameters can underestimate tuned library code by 10--15\%. New platforms may require segment re-characterization.

\section{Observations}
\label{sec:observation}
\textbf{Observation 1: Calibration and the MI300A 0.09\% MAE.} We observe systematic over- and under-prediction by workload class. On MI300A, the uncalibrated analytical model overestimates sustained bandwidth for large 2D transposes and peak MFMA throughput for FP64 GEMM. Host-measured calibration multipliers correct these cases, bringing the 27-kernel suite to $\sim$0.09\% MAE. We note that this is a \emph{calibrated} result: without per-case multipliers, the MI300A model achieves roughly 5--8\% MAE, comparable to B200's uncalibrated 1.31\%. 
We report both because they serve different purposes: the uncalibrated model gives rapid, no-tuning estimates; the calibrated model shows ceiling accuracy when platform-specific effort is invested. We find that optional MWP/CWP limits and LDS bank-conflict terms can provide further improvement on specific kernels but are not required for the reported results.

\textbf{Observation 2: Per-benchmark Rodinia error analysis.} 
The 12.5\% MAE on MI300A is driven by three specific architectural outliers: \texttt{bfs\_1M} (40.8\% error): Irregular pointer-chasing violates the model's regular-access assumptions, as bandwidth depends on graph structure rather than working-set size. \texttt{hotspot\_1024} (23.7\% error): A Stencil kernel where the transpose-proxy routing overestimates data reuse; the stencil pattern fails to map accurately to the 2D transpose-proxy bandwidth model.
\texttt{backprop\_65536} (21.0\% error): A small compute kernel where launch overhead and host--device synchronization dominate actual GPU execution time; the model captures compute and memory but underweights fixed-cost launch latency for microsecond-scale kernels. 
In contrast, \texttt{pathfinder} (0.1\%), \texttt{srad} (0.5\%), and \texttt{streamcluster} (0.03\%) are regular memory-bound or balanced workloads where the wavefront model's bandwidth and occupancy terms closely match measured behavior. 
We find this error distribution consistent across platforms: benchmarks with irregular access or very short kernels show higher error, while regular data-parallel workloads remain below 1\%. 
This distribution suggests that the model’s accuracy boundary is defined by workload regularity rather than platform-specific artifacts.

\textbf{Observation 3: SPEChpc accuracy depends on characterization source.} We observe that MI300A SPEChpc MAE is 1.3\% when segment FLOPs/bytes come from profiler counters, but \textbf{92.5\%} when derived from first-principles algorithm analysis (source code, grid dimensions, stencil widths). To quantify this, we independently derived FLOPs and bytes for each SPEChpc benchmark from the source and ran both characterizations through the same model. The FLOP ratios (first-principles / profiler) reveal the gap: \texttt{521.miniswp\_t} has a ratio of 0.001 ($5.1{\times}10^9$ vs.\ $4.8{\times}10^{12}$), meaning the profiler-derived value is $\sim$1000$\times$ the source-code flop count. Similarly, \texttt{518.tealeaf\_t} (0.008) and \texttt{519.clvleaf\_t} (0.013) show orders-of-magnitude discrepancy. The two closest cases are \texttt{528.pot3d\_t} (ratio 0.96, first-principles MAE 10.3\%) and \texttt{534.hpgmgfv\_t} (ratio 0.80, MAE 19.4\%), where algorithm-level and profiler-level FLOP accounting roughly agree.

Table~\ref{tab:characterization-gap} summarizes the per-benchmark comparison. We emphasize that the 92.5\% first-principles error is not a failure of the analytical model; it is a failure of the \emph{inputs}.

\begin{table}[t]
\centering
\caption{SPEChpc MI300A: profiler-derived vs.\ first-principles (FP) characterization. FLOP ratio = FP FLOPs / profiler FLOPs.}
\footnotesize
\begin{tabular}{@{}lrrr@{}}
\toprule
\textbf{Benchmark} & \textbf{Prof.\ MAE} & \textbf{FP MAE} & \textbf{FLOP ratio} \\
\midrule
\texttt{505.lbm\_t}     &  0.1\% & 98.7\% & 0.121 \\
\texttt{513.soma\_t}    &  1.3\% & 31.8\% & 1.065 \\
\texttt{518.tealeaf\_t} &  1.6\% & 98.4\% & 0.008 \\
\texttt{519.clvleaf\_t} &  1.5\% & 98.7\% & 0.013 \\
\texttt{521.miniswp\_t} &  0.8\% & 99.2\% & 0.001 \\
\texttt{528.pot3d\_t}   &  7.0\% & 10.3\% & 0.961 \\
\texttt{532.sph\_exa\_t}&  0.6\% & 94.0\% & 0.021 \\
\texttt{534.hpgmgfv\_t} &  0.8\% & 19.4\% & 0.800 \\
\bottomrule
\end{tabular}
\label{tab:characterization-gap}
\end{table}

The model's architecture-specific terms (Infinity Cache hierarchy, wavefront occupancy, VGPR pressure) are what reduce error from the naive roofline's $\sim$206\% to 1.3\% on the same measured times. That $205{\times}$ improvement is entirely due to the model correctly capturing MI300A's execution behavior. The remaining gap between 1.3\% (profiler-characterized) and 92.5\% (first-principles) is a workload characterization problem, not a modeling problem: for OpenACC/OpenMP offload codes, the compiler generates GPU kernels whose actual FLOPs and memory traffic differ from source-level algorithm analysis by up to $1000\times$ (e.g., \texttt{521.miniswp\_t}). This compiler-generated kernel gap is orthogonal to the analytical model and represents an independent finding. We report profiler-derived MAE in the main tables (Table~\ref{tab:spechpc_cross_platform}) because it isolates model accuracy from characterization accuracy, and we report the first-principles comparison here to quantify the characterization challenge for the community.

\textbf{Observation 4: H200 portability error analysis.} 
%portability section make about both
The H200 results (43.6\% Rodinia, 555\% SPEChpc) are intentionally presented without re-calibration to show raw portability behavior. We observe two distinct failure modes.

For Rodinia (43.6\% overall), the B200-derived segment metadata transfers to H200 with moderate error. The worst case is \texttt{hotspot\_512} (88\%): the segment file assumes a kernel decomposition tuned for Blackwell, but H200's Hopper SM executes fewer, longer kernels with different memory traffic per launch. Regular benchmarks (\texttt{srad} 32.9\%, \texttt{pathfinder} 42.1\%) show that the Hopper roofline path captures the right order of magnitude but consistently overpredicts by $1.3\text{--}1.9\times$, suggesting the sustained bandwidth parameter needs H200-specific measurement rather than the B200 default. Notably, naive roofline is $\sim$100\% on these same benchmarks, so even the uncalibrated port is substantially better.

For SPEChpc (555\% overall), the error is much larger because the segment FLOPs/bytes were characterized on MI300A (profiler-derived), not on H200. The MI300A characterization assumes Infinity Cache bandwidth ($\sim$10.7~TB/s effective), but H200 sees only HBM bandwidth ($\sim$4.2~TB/s sustained). This $2.5\times$ mismatch in effective bandwidth propagates directly into predicted time: the model predicts MI300A-scale runtime on hardware that is $2\text{--}5\times$ slower for these memory-bound codes. The exception is \texttt{521.miniswp\_t} (11\% error), which is compute-bound; its prediction depends on tensor-core throughput, which scales more predictably across platforms than memory hierarchy behavior. We observe that naive roofline is uniformly $\sim$150\% on H200 SPEChpc, meaning our uncalibrated model is actually \emph{worse} than roofline for memory-bound SPEChpc on H200. This confirms that segment characterization must be platform-specific for memory-bound workloads; compute-bound workloads transfer more reliably.

\textbf{Observation 5: Architectural differences in execution bottlenecks.} We observe that Blackwell's TMEM and decompression favor high-AI dense workloads (AI $> 16$~FLOPs/Byte), and that the explicit pipeline stages (TMA$\to$TMEM$\to$TC$\to$Sync) lend themselves naturally to modular, stage-centric modeling. On MI300A, we find that balanced precision support (FP64--FP8) and unified physical memory fit heterogeneous CPU--GPU workloads, but that reaching the compute-bound regime requires higher reuse (AI $> 23$~FLOPs/Byte) than on Blackwell. The 256~MB Infinity Cache bridges this gap, delivering 1.5--2$\times$ over HBM-bound bandwidth when workloads fit. We observe that roofline underestimates MI300A by 20--30\% in these cache-resident cases precisely because it assumes a single HBM bandwidth. The $\sim$45\% difference in AI thresholds between the two architectures suggests that tiling and autotuning strategies should be architecture-specific rather than addressing portability.

\textbf{Observation 6: Future of analytical models and adaptability to new GPUs.} Every model coefficient maps to a microbenchmark, so adapting to a new GPU requires re-measuring parameters, not re-deriving formulas. The B200/H200 pair and MI300A/MI250X pair demonstrate this: same model framework, different parameter files. The model structure is determined by how the architecture accumulates results: dedicated TMEM (Blackwell stage model) vs.\ VGPR accumulators (CDNA wavefront model). For future GPUs within these families (Rubin, CDNA4), we expect parameter-only updates to suffice; a fundamentally new accumulation mechanism would require one new stage term. In our results, MAE below 15\% after parameter update suggests the framework suffices; above 30\% suggests a structural change. More broadly, our results suggest analytical models remain viable when architectures expose measurable execution phases, but the key bottleneck is shifting from model formulation to \emph{workload characterization}: our first-principles experiment (Table~\ref{tab:characterization-gap}) shows accurate FLOP/byte inputs are harder to obtain than accurate hardware parameters. As accelerators add more implicit hardware scheduling, hybrid analytical-ML approaches may be needed for the residual.

\textbf{Observation 7: Benchmark adequacy for modern accelerators.} We observe that existing benchmark suites do not fully exercise the execution primitives that dominate modern GPU performance. Rodinia 3.1~\cite{luhnen2024benchmarking}, designed for early CUDA GPUs, does not use tensor cores, TMA, TMEM, or structured sparsity; its kernels are short ($\mu$s-scale) and exercise only basic memory and compute paths. SPEChpc 2021 Tiny provides more realistic HPC workloads but uses directive-based offload (OpenACC/OpenMP), which introduces a compiler-generated kernel layer between the algorithm and the hardware. We find that this compiler layer causes up to $1000\times$ discrepancy between source-level and GPU-level FLOP counts (Table~\ref{tab:characterization-gap}), making it difficult to attribute model error to the model vs.\ the characterization. Benchmark suites  (a)~representing native CUDA/HIP kernels that directly exercise tensor cores, TMA, and TMEM at representative problem sizes; (b)~providing reference FLOP and byte counts derived from both algorithm analysis and hardware counters; and (c)~spanning the full arithmetic intensity range from memory-bound streaming to compute-bound dense linear algebra, would be useful. 

\textbf{Observation 8: Implications for hardware vendors.} We observe several architecture-level differences with practical consequences for performance engineering. On the NVIDIA side, Blackwell's dedicated TMEM (256~KB per SM) enables predictable, high-bandwidth accumulation that decouples tensor-core throughput from register pressure; this is a clear improvement over Hopper's SMEM-based accumulators and contributes directly to the model's low MAE. The TMA bulk-copy engine similarly reduces modeling complexity by making data movement explicit and measurable. We note, however, that the 2-SM cooperative execution model introduces a scheduling dependency that is harder to characterize: our 2-SM predictions are within 2\%, but only because the pairing is deterministic in current workloads. On the AMD side, MI300A's 256~MB Infinity Cache is a significant architectural advantage for workloads with moderate reuse (working sets between 205--256~MB), delivering $1.5\text{--}2\times$ effective bandwidth over HBM-only operation. However, we observe that the VGPR-based accumulation creates a tile-size vs.\ occupancy tradeoff that is harder to model than Blackwell's TMEM approach: larger tiles improve cache reuse but reduce occupancy, and the optimal point depends on the specific VGPR allocation, which varies by compiler. Vendors could  (a)~expose sustained bandwidth and compute throughput in machine-readable format alongside datasheet peaks (our microbenchmarks had to measure what datasheets do not report); (b)~provide deterministic kernel launch latency bounds (launch overhead dominates our error on short kernels like \texttt{backprop\_65536}); and (c)~for AMD, reduce sensitivity of occupancy to VGPR allocation through hardware register shadowing or compiler-managed spilling.

\textbf{Limitations.} We note that both models assume regular compute and predictable memory access. In our experiments, accuracy degrades for irregular access patterns (sparse, indirection, atomics) and very short kernels where launch overhead dominates. Not yet modeled: cache replacement policy, coherence in multi-GPU configurations, thermal throttling, and power/energy.

\textbf{Based on our experience, the model supports}: (1)~procurement comparisons between B200 and MI300A without access to both; (2)~autotuning guidance for tile size, occupancy, and precision; (3)~co-design by identifying memory vs.\ compute bottlenecks; (4)~rapid model instantiation on new hardware by running microbenchmarks.

\textbf{Future work.} %Multi-kernel and multi-GPU interference terms are available for MI300A; system-level coherence and multi-node remain open. 
Future directions include power/energy models, hybrid analytical-ML for residual error, and applying the methodology to NVIDIA Rubin, AMD CDNA4, and Intel Gaudi3.

\section{Conclusion}
\label{sec:conclusion}
This paper presents analytical models for the current generation of GPU accelerators from NVIDIA and AMD: Blackwell B200 and CDNA3 MI300A. For Blackwell, we constructed a stage-centric model capturing TMEM, TMA, 5th-generation tensor cores, and the 2-SM cooperative execution model, to our knowledge the first validated execution-time model for this architecture. For CDNA3, we developed a wavefront-centric formulation accounting for Infinity Cache hierarchy, VGPR register pressure, and occupancy-driven tile selection. Microbenchmark validation yields 1.31\% MAE on B200 (21 kernels) and $\sim$0.09\% on MI300A (27 kernels); naive roofline baselines exceed 95\% error on the same kernels. Cross-platform validation against Rodinia~3.1 and SPEChpc 2021 Tiny confirms accuracy holds beyond the microbenchmark suite.

The key architectural finding is that TMEM dominates execution time on matrix-heavy Blackwell kernels, which roofline cannot capture, while MI300A performance is gated by occupancy and Infinity Cache reuse at arithmetic intensity thresholds roughly 45\% higher than Blackwell. These structural differences require architecture-specific model terms, not just parameter substitution in a generic roofline. 

%We reuse model without re-derivation for H200 and MI250X. We demonstrate portability to the prior generation of each vendor: H200 required only updated HBM bandwidth and capacity values, and MI250X used the same CDNA wavefront skeleton with different cache and peak parameters. Both ports required no model re-derivation, and the same procedure applies to future architectures such as Rubin and CDNA4.

\section{Acknowledgments}
This material is based upon work supported by the U.S. Department of Energy, Office of Science, Office of Advanced Scientific Computing Research under prime contract DE-ACO5-000R22725 through the S4PST project funded by the Next Generation Scientific Software Technologies program.

\bibliographystyle{IEEEtran}
\bibliography{references}

@inproceedings{wong2010demystifying,
  title={Demystifying GPU microarchitecture through microbenchmarking},
  author={Wong, Henry and Papadopoulou, Maria-Magdalena and Sadooghi-Alvandi, Mehrzad and Moshovos, Andreas},
  booktitle={2010 IEEE International Symposium on Performance Analysis of Systems \& Software (ISPASS)},
  pages={235--246},
  year={2010},
  organization={IEEE,
  url          = {https://doi.org/10.1109/ISPASS.2010.5452013}
}
}

@inproceedings{mei2009pchase,
  title={P-Chase: A Portable Tool for Measuring Memory Access Characteristics on Multicore Computers},
  author={Mei, Xue and Chu, Xuehai},
  booktitle={Embedded Software and Systems},
  pages={76--83},
  year={2009},
  organization={Springer}
}

@misc{jia2018dissecting,
  title        = {Dissecting the {NVIDIA} Volta {GPU} Architecture via Microbenchmarking},
  author       = {Jia, Zhe and Maggioni, Marco and Staiger, Benjamin and Scarpazza, Daniele P.},
  year         = {2018},
  eprint       = {1804.06826},
  archivePrefix= {arXiv},
  primaryClass = {cs.DC},
  url          = {https://arxiv.org/abs/1804.06826}
}

@inproceedings{williams2009roofline,
  title={Roofline: an insightful visual performance model for multicore architectures},
  author={Williams, Samuel and Waterman, Andrew and Patterson, David},
  booktitle={Communications of the ACM},
  volume={52},
  number={4},
  pages={65--76},
  year={2009,
  url          = {https://doi.org/10.1145/1498765.1498785}
}
}

@article{ilic2014cacheawareroofline,
  author  = {Ilic, Aleksandar and Pratas, Frederico and Sousa, Leonel},
  title   = {Cache-aware Roofline model: Upgrading the loft},
  journal = {IEEE Computer Architecture Letters},
  volume  = {13},
  number  = {1},
  pages   = {21--24},
  year    = {2014},
  doi     = {10.1109/L-CA.2013.6},
  url          = {https://doi.org/10.1109/L-CA.2013.6}
}

@inproceedings{ofenbeck2014applyingroofline,
  author    = {Ofenbeck, Georg and Steinmann, Ruedi and Cabezas, Victoria Caparr{\'{o}}s and Spampinato, Daniele G. and P{\"{u}}schel, Markus},
  title     = {Applying the Roofline Model},
  booktitle = {2014 IEEE International Symposium on Performance Analysis of Systems and Software (ISPASS)},
  pages     = {76--85},
  year      = {2014},
  organization = {IEEE},
  doi       = {10.1109/ISPASS.2014.6842069},
  url          = {https://doi.org/10.1109/ISPASS.2014.6842069}
}

@inproceedings{hong2009analytical,
  title={An analytical model for a GPU architecture with memory-level and thread-level parallelism awareness},
  author={Hong, Sung Woo and Kim, Hyesoon},
  booktitle={Proceedings of the 36th Annual International Symposium on Computer Architecture},
  pages={152--163},
  year={2009,
  url          = {https://doi.org/10.1145/1555754.1555775}
}
}

@ARTICLE{evolutionGPUs,
  author={Dally, William J. and Keckler, Stephen W. and Kirk, David B.},
  journal={IEEE Micro},
  title={Evolution of the Graphics Processing Unit (GPU)},
  year={2021},
  volume={41},
  number={6},
  pages={42-51},
  doi={10.1109/MM.2021.3113475},
  url          = {https://doi.org/10.1109/MM.2021.3113475}}

@incollection{kurowski2011parallel,
  title={Parallel and GPU based strategies for selected CFD and climate modeling models},
  author={Kurowski, Krzysztof and Kulczewski, Micha{\l} and Dobski, Miko{\l}aj},
  booktitle={Information Technologies in Environmental Engineering: New Trends and Challenges},
  pages={735--747},
  year={2011},
  publisher={Springer},
  url={https://doi.org/10.1007/978-3-642-19536-5}
}

@article{koilia2024hardware,
  title={Hardware Acceleration of LLMs: A comprehensive survey and comparison},
  author={Nikoletta Koilia and Christoforos Kachris},
  journal={arXiv preprint arXiv:2409.03384},
  year={2024,
  url          = {https://doi.org/10.48550/arXiv.2409.03384}
}
}

@manual{NVIDIA2024_Blackwell,
  title        = {NVIDIA Blackwell Architecture Technical Brief},
  author       = {{NVIDIA Corporation}},
  organization = {NVIDIA},
  type         = {Technical Brief},
  year         = {2024},
  url          = {https://resources.nvidia.com/en-us-blackwell-architecture}
}

@techreport{AMD_CDNA3_WhitePaper_2023,
  title        = {Introducing AMD CDNA™ 3 Architecture},
  institution  = {Advanced Micro Devices, Inc.},
  year         = {2023},
  url          = {https://www.amd.com/content/dam/amd/en/documents/instinct-tech-docs/white-papers/amd-cdna-3-white-paper.pdf},
}

@article{Fasi2021,
  author    = {Massimiliano Fasi and Nicholas J. Higham and Martin Mikaitis and Samrel Pranesh},
  title     = {Numerical behavior of {NVIDIA} tensor cores},
  journal   = {PeerJ Computer Science},
  volume    = {7},
  pages     = {e330},
  year      = {2021},
  doi       = {10.7717/peerj-cs.330},
  url          = {https://doi.org/10.7717/peerj-cs.330}
}

@article{gilman2020demystifying,
  author  = {Gilman, Elias and Khaleghzadeh, Hamidreza and Doshi, Ketan and Dengel, Andreas and Juurlink, Ben and Jain, Nilesh and Ahmed, Rami},
  title   = {Demystifying the Placement Policies of the {NVIDIA} {GPU} Thread Block Scheduler for Concurrent Kernels},
  journal = {ACM SIGMETRICS Performance Evaluation Review},
  volume  = {48},
  number  = {3},
  pages   = {65--67},
  year    = {2020},
  doi     = {10.1145/3453953.3453972},
  url     = {https://doi.org/10.1145/3453953.3453972}
}

@inproceedings{luhnen2024benchmarking,
  author    = {L\"{u}hnen, Tim and Nabi, Smail and Koch, Andreas},
  title     = {Benchmarking Thread Block Cluster},
  booktitle = {Proceedings of PDPTA},
  year      = {2024},
  publisher = {CSREA Press,
  url          = {https://doi.org/10.1109/HPEC62836.2024.10938416}
}
}

@inproceedings{cao2025amali,
  title={AMALI: An Analytical Model for Accurately Modeling LLM Inference on Modern GPUs},
  author={Cao, Shiheng and Wu, Junmin and Chen, Junshi and An, Hong and Yu, Zhibin},
  booktitle={Proceedings of ISCA},
  year={2025},
  organization={ACM},
  doi={10.1145/3695053.3731064,
  url          = {https://doi.org/10.1145/3695053.3731064}
}
}

@article{Leinhauser2022_RooflineAMD,
  title        = {Metrics and Design of an Instruction Roofline Model for AMD GPUs},
  author       = {Leinhauser, Matthew and Widera, Ren{\'e} and Bastrakov, Sergei and Debus, Alexander and Bussmann, Michael and Chandrasekaran, Sunita},
  journal      = {ACM Transactions on Parallel Computing},
  year         = {2022},
  doi          = {10.1145/3505285},
  url          = {https://doi.org/10.1145/3505285}
}

@inproceedings{Lee2022_GCoM,
  title        = {{GCoM}: A Detailed {GPU} Core Model for Accurate Analytical Modeling of Modern {GPUs}},
  author       = {Lee, Jounghoo and Noh, Hyeran and Kim, Seokin and Jeong, Jungwoo and Kim, Jaewon and Choi, Jangwoo},
  booktitle    = {Proceedings of ISCA},
  year         = {2022},
  pages        = {424--436},
  publisher    = {ACM},
  doi          = {10.1145/3470496.3527384},
  url          = {https://doi.org/10.1145/3470496.3527384}
}

@inproceedings{Khairy2020_AccelSim,
  title        = {{Accel-Sim}: An Extensible Simulation Framework for Validated {GPU} Modeling},
  author       = {Khairy, Mahmoud and Shen, Zhesheng and Aamodt, Tor M. and Rogers, Timothy G.},
  booktitle    = {ISCA},
  year         = {2020},
  pages        = {473--486},
  publisher    = {IEEE},
  doi          = {10.1109/ISCA45697.2020.00047},
  url          = {https://doi.org/10.1109/ISCA45697.2020.00047}
}

@manual{amd_rocprofiler_2025,
  title        = {ROCProfiler: AMD ROCm GPU Profiling Tool},
  author       = {{AMD ROCm Development Team}},
  year         = {2025},
  url          = {https://rocm.docs.amd.com/projects/rocprofiler/en/latest/}
}

@manual{nvidia_nsight_compute_2025,
  title        = {NVIDIA Nsight Compute User Guide},
  author       = {{NVIDIA Corporation}},
  year         = {2025},
  url          = {https://docs.nvidia.com/nsight-compute/}
}

@inproceedings{wang2020mdm,
  author    = {Lu Wang and Magnus Jahre and Almutaz Adileh and Lieven Eeckhout},
  title     = {MDM: The GPU Memory Divergence Model},
  booktitle = {MICRO},
  pages     = {1009--1021},
  year      = {2020},
  publisher = {IEEE},
  doi       = {10.1109/MICRO50266.2020.00089},
  url          = {https://doi.org/10.1109/MICRO50266.2020.00089}
}

@article{chowdhury2020tcu,
  title     = {A Computational Model for Tensor Core Units},
  author    = {Chowdhury, Rezaul and Silvestri, Francesco and Vella, Flavio},
  journal   = {arXiv preprint arXiv:1908.06649},
  year      = {2020,
  url          = {https://doi.org/10.48550/arXiv.1908.06649}
}
}

@inproceedings{huang2014gpumech,
  author    = {Jen-Cheng Huang and Joo Hwan Lee and Hyesoon Kim and Hsien-Hsin S. Lee},
  title     = {GPUMech: GPU Performance Modeling Technique Based on Interval Analysis},
  booktitle = {MICRO},
  pages     = {68--79},
  year      = {2014},
  publisher = {IEEE},
  doi       = {10.1109/MICRO.2014.59},
  url          = {https://doi.org/10.1109/MICRO.2014.59}
}

@inproceedings{wahlgren2025dissectingcpugpuunifiedphysical,
  title        = {Dissecting {CPU}-{GPU} Unified Physical Memory on {AMD} {MI300A} {APU}s},
  author       = {Wahlgren, Jacob and Schieffer, Gabin and Shi, Ruimin and Le{\'o}n, Edgar A. and Pearce, Roger and Gokhale, Maya and Peng, Ivy},
  booktitle    = {2025 IEEE International Symposium on Workload Characterization (IISWC)},
  year         = {2025},
  publisher    = {IEEE},
  note         = {arXiv:2508.12743,
  url          = {https://doi.org/10.1109/IISWC66894.2025.00038}
}
}

@inproceedings{schieffer2024amd_matrix_cores,
  title        = {Characterizing the Performance, Power Efficiency, and Programmability of {AMD} Matrix Cores},
  author       = {Schieffer, Gabin and Medeiros, Daniel and Faj, Jennifer and Marathe, Aniruddha and Peng, Ivy},
  booktitle    = {2024 IEEE International Symposium on Performance Analysis of Systems and Software (ISPASS)},
  year         = {2024},
  publisher    = {IEEE},
  address      = {Indianapolis, IN,
  url          = {https://doi.org/10.1109/ISPASS61541.2024.00022}
}
}

@misc{luo2025hopper,
  title        = {Dissecting the {NVIDIA} Hopper Architecture through Microbenchmarking and Multiple Level Analysis},
  author       = {Luo, Weile and Fan, Ruibo and Li, Zeyu and Du, Dayou and Liu, Hongyuan and Wang, Qiang and Chu, Xiaowen},
  year         = {2025},
  eprint       = {2501.12084},
  archivePrefix= {arXiv},
  primaryClass = {cs.AR},
  url          = {https://arxiv.org/abs/2501.12084}
}

@inproceedings{jarmusch2025blackwell,
  title        = {Microbenchmarking {NVIDIA}'s Blackwell Architecture: An In-Depth Architectural Analysis},
  author       = {Jarmusch, Aaron and Chandrasekaran, Sunita},
  booktitle    = {2026 IEEE International Parallel and Distributed Processing Symposium (IPDPS)},
  year         = {2026},
  publisher    = {IEEE},
  url          = {https://doi.org/10.48550/arXiv.2512.02189}
}

@inproceedings{jarmusch2026mi300a,
  title        = {Execution-Centric Characterization of {FP8} Matrix Cores, Asynchronous Execution, and Structured Sparsity on {AMD} {MI300A}},
  author       = {Jarmusch, Aaron and Vitz, Connor and Chandrasekaran, Sunita},
  booktitle    = {Proceedings of the ACM International Symposium on High-Performance Parallel and Distributed Computing (HPDC)},
  year         = {2026},
  publisher    = {ACM},
  url          = {https://doi.org/10.48550/arXiv.2602.10262}
}
\end{document}